\documentstyle[11pt]{article}
\input{epsf}

\title{Slipher and the Nature of the Nebulae}
\author{Ken Freeman \\ The Australian National University}
\date{}

\begin{document}
\maketitle
\def\eg{{\rm e.g. }}
\def\ie{{\rm i.e. }}
\def\etal{{\rm et~al. }}

\abstract { Why do some discoveries, which appear in hindsight to be obviously major discoveries, have so little impact when they were made.   These are discoveries that were ahead of their time, but for some reason the scientific community was not ready to absorb them. I got interested in Slipher after learning that by 1914 he had observed nebular redshifts up to 1000 km s$^{-1}$.  Why did this not convince people at the time that the nebulae were extragalactic ?  Was this another example, like Zwicky's (1933) discovery of dark matter in the Coma cluster,  of a discovery which was too far ahead of its time ?  I conclude that Slipher's situation was different from Zwicky's. The significance of Slipher's nebular redshifts was partly recognised  at the time, but I believe that its full significance was masked by van Maanen's work which turned out to be erroneous and reduced the impact of Slipher's discoveries. }

\section{Introduction}

I am curious to know why some discoveries, which we now see as very important, made relatively little impact at the time when they were made. This issue is related (but not one to one) to why some people  don't get the credit for their discoveries. I will start this talk with two major discoveries related to dark matter in galaxies which got little response from the community at the time  

\begin{itemize}
\item Zwicky and the mass of the Coma cluster (1933, 1937)
\item Kahn \& Woltjer (1959) on the timing mass for the M31-Galaxy system
\end{itemize}
and then contrast these discoveries with the later discoveries of pulsars (1967) and dark matter in dwarf spheroidal galaxies (1983), both of which had an immediate impact.  Then I will turn to Slipher's work on galaxy redshifts.

\section{Zwicky's discovery of dark matter in the Coma cluster}
In 1933, Zwicky reported his measurement of the velocity dispersion of galaxies in the Coma cluster of galaxies.  Using virial theorem techniques that are used today, he showed that the mass of the cluster is much larger than the likely sum of the masses of the individual galaxies.  A similar result was found  by  Smith (1936) for the Virgo cluster.  Zwicky's value for the velocity dispersion of Coma is close to present estimates. This profound discovery got little response, other than Smith's follow-up study for Virgo.  

It took another 35 years for the saga of dark matter (DM) in galaxies to take off.   And then it took another 20 years for us to learn that clusters of galaxies have the universal baryon content (about 16\% of their mass, including the hot gas which dominates the baryon component of the cluster), and that the rest of their mass is in the form of DM which is believed to reside partly in the galaxies and partly in the cluster itself  (White et al 1993).  

Why did Zwicky's important and apparently straightforward  discovery get so little response ? Did the community already regard Zwicky with suspicion  (or did that come later) ?  Or was it due to suspicion of results for which there is no existing theoretical framework ?  Some would agree with this irritating quote from Eddington (1947): ``It is \dots\ a good rule not to put overmuch confidence in observational results until they are confirmed by theory".   (I am grateful to Matt Stanley for providing the source of this quote.)  Why didn't astronomers follow up Zwicky's important discovery with spectroscopy of nearby groups of galaxies like the M81 and Leo groups, which would be much easier to study than the more distant galaxies in the Coma cluster?

Zwicky does now get full credit for his discovery.   This is a partial counter-example to Stigler's Law of Eponymy, that no scientific discovery is named after its original discoverer.  Big new ideas only tend to catch on when the scientific community is ready for them.  

\section{Kahn \& Woltjer (1959)}
The Kahn \& Woltjer (1959) paper is about the motion of M31 relative to the Milky Way.  This is another important paper that did not have the impact it deserved. M31 is about 750 kpc from the Milky Way and approaches the Milky Way at 118 km s$^{-1}$.   Kahn \& Woltjer noted that the two galaxies are now approaching each other. They assumed that (i) the two galaxies were formed close together, (ii)  that their combined mass was sufficient to make them a bound unit, and (iii) that they have performed the larger part of at least one orbit with a period of no more than 15 Gyr.  They adopted a radial orbit (which turned out to be very close to the truth) and simple Keplerian dynamics to show that the mass of the (M31--Milky Way) system is about 20 times larger than the likely masses of the stars of the two galaxies. Their distance scale and stellar masses were roughly right.  Their estimate of the total mass was $ > 2.10^{12}$ M$_\odot$  (present estimates are about $3.10^{12}$  M$_\odot$: van der Marel et al. 2012).  

Kahn \& Woltjer excluded the possibility of intergalactic stars and argued that the extra required matter was in the form of intergalactic ionized gas.  Their paper went on to explore the distorting effects of this gas on the Galactic disk. Maybe that was a distraction.  I remain surprised that this paper had so little impact. The argument is simple and correct and has survived to the present. Kahn and Woltjer were both very respectable and well-regarded researchers.  Their work should have started the dark matter revival in 1959.

My guess is that the situation for the Kahn \& Woltjer paper was much as for Zwicky's much earlier paper. There was simply no theoretical framework within which to interpret the observation of such a large total mass for the (M31--Milky Way) system.  Their argument is direct and compelling but made almost no impact at the time.  Some weak contrary evidence (e.g. Godfredsen 1961) based on virial theorem arguments, that concluded that extra mass was not needed to bind the Local Group, may have provided a welcome escape.  
I will speculate later that Slipher's work suffered a somewhat parallel situation. His discoveries did not have the impact they deserved in settling the controversy about the nature of the nebulae because of work by van Maanen which convinced much of the community at the time but turned out to be incorrect.
 
When the dark matter revival came in the 1970s, it was based observationally on the 21-cm rotation curves of spiral galaxies, which showed that they have massive and extended dark halos.  This argument was not much more direct than the Kahn \& Woltjer argument but it made an impact, even though the evidence for dark matter from rotation curves was not really secure until about 1978 when the high quality 21-cm radio synthesis data became available. The difference was that, by the early 1970s, there was a theoretical framework on the need for dark halos to stabilize disks against bar formation (e.g. Ostriker \& Peebles 1973).  This framework helped to sustain the quest for dark matter in galaxies through the period of controversy in the 1970s.  People appear to be ready to believe observations that fit into some theoretical framework, even if the observations have a sounder basis than the theory. It allows them to come to grips with startling observations, as Eddington pointed out.  In this case, the sustaining theory turned out to be only partly relevant.   Most disks are now known to have bars, so there is no need to invoke a dark halo to stabilize them.  In fact, dark halos are now known to be needed to promote the growth of bars via angular momentum transport.

\section{Two discoveries that had an immediate impact} 
The discovery of pulsars in 1968 by Hewish, Bell et al. was unexpected and  had an immediate impact.  It was followed by a quick stream of papers : after a brief ``little green men'' interlude\footnote{My recollection from that time is that the discoverers  briefly considered the possibility that they had detected a signal from an extraterrestrial civilization, but soon discarded this possibility as further pulsars were discovered.},  the focus quickly moved to the current spinning neutron star explanation.   Why did it get such a quick response ?   Probably because the basic theoretical framework was already in place, from Baade \& Zwicky (1934) and later theoretical developments:  people quickly made the connection.

Another example is the discovery of the high fractions of dark matter in dwarf spheroidal (dSph) galaxies,  starting with the work of Faber \& Lin (1983) and Aaronson (1983).  The arguments were not strong at the time and the discovery came as a surprise,  but there was only limited scepticism.  Dark matter in dSph galaxies remains a very active field today. Current estimates of their mass-to-light ratios exceed 1000 solar units (e.g. Wilkinson et al. 2006) for some of the faintest dSph galaxies. Why did this discovery get such a quick response ?  Probably because the basic observational infrastructure of dark matter in brighter galaxies was already there, plus the theoretical ideas on the role of dark matter in galaxy formation (e.g. White \& Rees 1978) .

\section{Slipher's contribution to nebular spectroscopy}

After coming to Lowell Observatory, Slipher developed expertise in spectroscopy.  In 1904, he wrote a paper on the Lowell spectrograph,  built by the Brashear company for planetary spectroscopy, to go on the 61 cm Clark refractor. Lowell requested Slipher to acquire spectra of  nebulae, because Lowell believed that the nebulae may be solar systems in the process of formation.  Nebular spectroscopy was a major observational  challenge because of their low surface brightness, and  Slipher  modified the Brashear spectrograph for nebular spectroscopy. His modifications made it possible to acquire spectra of the nebulae and measure their redshifts and detect rotation (see talks by Smith and by Thompson at this meeting).

With the original optics, used for stellar and planetary spectroscopy, the linear dispersion of the spectrograph was about 11 \AA\ mm$^{-1}$ corresponding to a resolving power of about 22,000 (this is typical of the spectrographs of the time and is on the low side of high resolution spectroscopy today).  After modification for nebular spectroscopy, the linear dispersion of the spectrograph was 140 tenth-meters per mm (140 \AA\ mm$^{-1}$) which is  more than adequate to measure nebular velocities and detect their rotation. The nebular exposures were long: 20 to 40 hours.

For about a decade, Slipher provided most of the nebular velocities.  He mentions confirmations by Wright (Lick), Wolf (Heidelberg) and Pease (Mt Wilson).   Fath (Lick) and others had already acquired nebular spectra but, for technical reasons to do with wavelength calibration, had not measured their radial velocities.   

Here are Slipher's most significant papers on nebular spectroscopy:
\begin{itemize}
\item  the 1913 paper on the radial velocity of M31.  This appears to be the first measurement of the radial velocity of a spiral galaxy.  Its velocity is about $-300$ km s$^{-1}$.  Slipher comments:  ``That the velocity of the first spiral observed should be so high intimates that the spirals as a class have higher velocities than do the stars and that it might not be fruitless to observe some of the more promising spirals for proper motion''.
(It was the observations of proper motion within spirals, later shown to be incorrect, that may have cost Slipher the credit for discovering the extragalactic nature of the nebulae.)
\item in 1914, Slipher published his detection of rotation in the Sombrero galaxy NGC 4594.  It has a high surface brightness bulge with a rotational velocity of about 200 km s$^{-1}$ and a radial velocity of about 1000 km s$^{-1}$.
\item Slipher's 1915 paper gave radial velocities for 15 spirals: 13 are positive and go up to 1100 km s$^{-1}$.
The mean velocity is about 400 km s$^{-1}$, about 25 times the average velocity of the Galactic stars.
\item by 1917, Slipher had velocities of 25 spirals, all positive except for some Local Group galaxies and M81.  The mean velocity of the nebulae was now about 30 times the average velocity of the Galactic stars.
\end{itemize}

How accurate are Slipher's velocities relative to modern values ?  The dispersion of Slipher's velocities about the modern values is 112 km s$^{-1}$, close to his own estimate of the uncertainty.  

The velocities of stars were known from the work of Boss, Campbell, Kapteyn and others to increase with spectral type, from about 6 km s$^{-1}$ at B to 15 km s$^{-1}$ at K and then to 27 km s$^{-1}$ for the planetary nebulae  (Smith 2008).  (This effect is still not entirely understood.)  Could the nebulae be Galactic objects, much further up this evolutionary chain ?   By 1916 there were already some ideas about obscuring material in the Milky Way.   It was long known that the nebulae avoided the Galactic plane, which again favored an interpretation putting them outside the Milky Way.  Further support came from observations of extragalactic novae.

Slipher was well aware of the significance of his observations.  The velocities of the nebulae are much larger than those of the Galactic stars.  He inferred that they lie outside the Milky Way. In his 1917 paper, he wrote:
\begin{quote}  It has for a long time been suggested that the spiral nebulae are stellar systems seen at great distances.  This is the so-called ``island universe'' theory, which regards our stellar system and the Milky Way as a great spiral nebula which we see from within.  This theory, it seems to me, gains favor in the present observations.
\end{quote}

\noindent A letter from Hertzsprung to Slipher in 1914 makes the same point.  Hertzsprung wrote:
\begin{quote}
My \dots\ congratulations to your beautiful discovery of the great radial velocity of some spiral nebulae.
It seems to me, that with this discovery the great question, if the spirals belong to the system of the milky way or not, is answered with great certainty to the end, that they do not.
\end{quote}

\noindent Although some (like J.H. Reynolds) had their doubts about the data, Slipher had a convincing response, with confirmation of the velocities from others.  Slipher's discoveries were well known at the time.   Why did they not settle the issue about the nature of the nebulae ?   

Van Maanen's work on proper motions in nearby galaxies confused the issue. He was a respected figure and his results were taken seriously: they were  not really discredited until about 1935 (see M. Way's paper in this volume).   Van Maanen's papers on proper motions began to appear in 1916.  His results may have been known to astronomers a year or two earlier, and it seems likely that his work was influencing the thinking at the time when Slipher was assembling his 1915 and 1917 catalogs of nebular velocities.

In the 1920 Shapley-Curtis debate,   Shapley argued that the nebulae are just nearby clouds and the universe is one big Galaxy. Curtis argued that the nebulae are galaxies like our own, far outside the Milky Way.  In the end, Hubble's 1925 cepheid study on M31 settled the question. I am interested in why the "great debate" ever took place, given Slipher's discoveries.  Hale appears to have been the prime mover in setting up this debate (see Smith 2008).  van Maanen's work would have been well known to the Mt Wilson astronomers, and it may be that Hale's view of the island universe controversy (and also Shapley's) was unduly influenced by van Maanen's results, in contrast to the  view of Curtis from Lick.

There is some similarity of this controversy and the dark matter controversy of the 1970s, but in reverse. The development of the dark matter story was supported by theory which we now believe to be only partly relevant, while Slipher's island universe story was delayed by erroneous observations.  Progress is not always linear. 

I began preparation of this talk, believing that Slipher's nebular velocities did not have an impact in their time, similar to the lack of impact of Zwicky's discovery of dark matter in the Coma cluster.  But that view is not correct:  Slipher's situation is not at all comparable with Zwicky's.  Slipher's work made an impact at the time, but his problem of recognition came later.  I speculate that, without the confusion from van Maanen's incorrect proper motions, the nature of the nebulae would have been clearer.  Maybe the Shapley-Curtis debate  would not have happened and Slipher would have got the credit for identifying the nebulae as extragalactic, which I think he deserves.

\section{Hubble and the discoveries of others}

Hubble's name comes up frequently in discussions about  the appropriation of discoveries by others (see M. Way's talk at this meeting) and  it is interesting to ask why.  I am aware of at least  four examples involving Hubble:
\begin{enumerate}
\item At this meeting, we are concerned with the lack of recognition of Slipher's redshift achievements until long after the event. His work did have impact on the community at the time, but later it did not get the recognition it deserved in the context of Hubble's expanding universe.
\item The issue with the Hubble redshift-distance law and Lema\^itre's contribution has been much discussed recently.  Lema\^itre's discovery of the law a few years earlier was not recognized properly. A translation of Lema\^itre's paper into English omitted an observational section that described what later became known as the Hubble law and included Lema\^itre's derivation of the Hubble constant.  Mario Livio (2011) has argued convincingly that this omission was Lema\^itre's choice.
\item In a recent book (Block \& Freeman 2008), David Block and I identified two other incidents involving Hubble and the UK astronomer John Reynolds.   The Hubble classification of galaxies was basically invented by Reynolds.   Hubble knew about Reynold's work, and their correspondence about galaxy classification around 1919 is in the RAS archives.
\item  The other Hubble law. The same Reynolds discovered that the surface brightness distribution in elliptical galaxies can be represented by a simple law of the form  $I(R) = I_\circ(1 + R/a)^{-2}$ where $R$ is the projected radius on the sky and $a$ is a scale length.  This law became known as the Hubble distribution but more recently as the Hubble-Reynolds law.  
\end{enumerate}

There is a view that Hubble was not generous in acknowledging the contributions of others. Some who knew him regarded him poorly in this respect.  On the other hand, some of us are careless about picking up ideas and forgetting where they came from. It still happens.  Geography,  institutional rivalry and culture may also be significant elements in this behaviour.  For others, modesty is more important than credit.


\bigskip
\begin{thebibliography}{}
\bibitem{}Aaronson, M. 1983. ApJ, 266, L11 
\bibitem{}Baade, W. \& Zwicky, F. 1934. Proc. Nat. Acad. Sci, 20, 254 
\bibitem{}Block, D. \& Freeman, K. 2008. ``Shrouds of the Night''  (New York: Springer) 
\bibitem{} Eddington, A. 1947. ``New Pathways in Science", Cambridge University Press,  (reprint), page 21
\bibitem{}Faber, S. \& Lin, D. 1983. ApJ, 266, L17 
\bibitem{}Godfredsen, E.A. 1961. ApJ, 134, 257 
\bibitem{}Hewish, A., Bell, J et al. 1968.  Nature, 217, 709 
\bibitem{}Hubble, E. 1925. Observatory, 48, 139 
\bibitem{}Kahn, F. \& Woltjer, L. 1959. ApJ, 130, 705 
\bibitem{}Livio, M. 2011.  Nature, 479, 171 
\bibitem{}Ostriker, J.P. \& Peebles, P.J. 1973. ApJ, 186, 467 
\bibitem{}Slipher, V.M. 1904. ApJ, 20, 1 
\bibitem{}Slipher, V.M. 1913. Bull. Lowell Obs., No. 58 
\bibitem{}Slipher, V.M. 1914.  Bull. Lowell Obs., No. 62 
\bibitem{}Slipher, V.M. 1915.  Amer. Astron. Soc, 17th meeting (Pop. Astr., 23, 21) 
\bibitem{}Slipher, V.M. 1917.  Proc. Amer. Phil. Soc., 56, 403 
\bibitem{}Smith, R.  2008. JHA, 39, 91 
\bibitem{}Smith, S. 1936. ApJ, 83, 23 
\bibitem{}van der Marel, R., Fardal, M. et al. 2012. ApJ, 753, 8 
\bibitem{}White, S., Navarro, J. et al. 1993. Nature, 366, 429 
\bibitem{}White, S. \& Rees, M. 1978. MNRAS, 183,  341 
\bibitem{}Wilkinson, M., Kleyna, J. et al. 2006. ESO Messenger, 124, 25 
\bibitem{}Zwicky, F. 1933.  Helvetia Physica Acta, 6, 110 
\bibitem{}Zwicky, F.  1937. ApJ, 86, 217 
\end{thebibliography}
\end{document}